# Energy-efficient traffic engineering for future core networks

G. Athanasiou

A general problem formulation for energy-efficient traffic engineering for future core networks is presented. Moreover, a distributed heuristic algorithm that provides jointly load balancing and energy efficiency is proposed, approaching in this way the optimal network operation in terms of throughput and energy consumption.

*Introduction:* Traffic engineering (TE) plays a crucial role in determining the performance and reliability of network deployments. A major challenge in traffic engineering is how to cope with dynamic and unpredictable changes in traffic demands and how the network could handle possible traffic variations in a way that *load balancing*, *congestion avoidance* and *efficient service provisioning* are ensured. A more straightforward explanation of *TE* is given in [1]: *"to put the traffic where the network bandwidth is available"*. The rapid growing of the users and the services that must be supported, the spreading of broadband access in conjunction with the increased energy prices affected the demand for energy-aware service provisioning. A challenging task is to identify the main parts of the Internet that dominate its power consumption and investigate methods for improving energy consumption [2]. The first attempt to introduce energy savings in the Internet was made in [3]. The authors in [4] discuss the idea of dynamically turning part of the network operations into sleeping mode, during light utilization periods, in order to minimize the energy consumption. The authors in [5] identified the power saving problem in the Internet, and propose sleeping as the approach to conserve energy. Moreover, routing, rate adaptation and network control are mobilized towards energy-efficient network operation [6]. Unfortunately, none of these approaches provide a problem formulation in the direction of jointly studying the "traditional" objectives and the new objectives (energy-awareness) of *TE*. This work is an attempt to put these objectives under a joint problem formulation and to propose lightweight solutions that could be applied in real network deployments. We provide a joint problem formulation for optimal energy-aware load balancing in the network. Then, we propose a distributed *ENergy-Efficient TRaffic Engineering (ENTRE)* scheme that smoothly introduces the aforementioned major issues in real deployments.

*Energy-efficient traffic engineering approach:* We consider that multiple paths (MPLS tunnels) are used to deliver traffic from the ingress to the egress routers. Formally, we assume that for each ingress-egress node pair $i$ the traffic demand is $T_i$ and multiple paths $P_i$ could be used to deliver the traffic from the ingress to the egress node. Therefore, a fraction of the traffic $x_{ip}$ is routed along path $p$ ($p \in P_i$). In addition, the energy consumption of an active link is affected by the maximum rate that the link can support (10Mbps, 100Mbps, etc) and the current utilization of that link. The calculation of the energy consumption of link $l$, $e_l$, with capacity $c_l$, is based on the simple model proposed in [7] (used also in several approaches in literature):

$$e_l = PowerConsumption(c_l) \times UtilizationFactor(l)$$

*PowerConsumption($c_l$)* is the base power consumption of a link with capacity $c_l$ and *UtilizationFactor(l)* is the scaling factor to account for the utilization of link $l$. Table 1 contains the definition of the variables used in our problem formulation.

In our formulation, the optimal splitting of the traffic is performed with the objective: *Assure that the maximum link utilization (total traffic on an active link divided by the link capacity) in the network is minimized. In this way resource-efficient (in this case link utilization/bandwidth) and balanced/stable network operation is achieved* [8]. In order to introduce energy-awareness we raise a second objective: *Assure that the energy consumption of the active routes is balanced in the network. That is, find the route with maximum energy consumption and minimize it*. We combine the previous single objectives in a unified objective function that takes into account the link utilization and the energy consumption:

$$\min_{x_{ip}} \max_{l \in L_p} \sum_{i \in IE} \sum_{p \in P_i} \left( \frac{x_{ip} T_i}{c_l} \times \sum_{k \in L_p} e_k \right),$$

subject to:

$$x_{ip} \geq 0, \forall p \in P_i, \forall i \in IE$$
$$c_l \geq \sum_{i \in IE} \sum_{p \in P_l} x_{ip} T_i, \forall l \in L$$
$$\sum_{p \in P_i} x_{ip} = 1, \forall i \in IE$$
$$x_{ip} = [0,1], \forall p \in P_i, \forall i \in IE$$

The fraction of traffic for a specific *IE* node pair sent across a path cannot be negative and the capacity of each link cannot be outreached.

**Table 1:** Variables used in our problem formulation.

| Var. | Description |
|---|---|
| $L$ | Set of links in the network |
| $IE$ | Set of Ingress to Egress node pairs |
| $e_l, E_{ip}$ | Energy consumption of the port connected to link $l$, path $p$ |
| $P_i$ | Set of paths of Ingress to Egress node pair $i$ |
| $T_i$ | Traffic demand of Ingress to Egress node pair $i$ |
| $u_l, U_{ip}$ | Utilization of link $l$, path $p$ |
| $c_l$ | Capacity of link $l$ |
| $x_{ip}$ | Fraction of traffic of Ingress to Egress node pair $i$, sent through the path $p$ |
| $r_{ip}$ | Traffic of Ingress to Egress node pair $i$, sent through path $p$ |
| $P_l$ | Set of paths that go through link $l$ |
| $L_i$ | Set of links that are crossed by the set of paths $P_i$ |
| $B_p t_m$ | Number of bits sent along the path $p$ during $t_m$ seconds |
| $T_E, T_U$ | Thresholds related to utilization and energy consumption |

We now present an *ENergy-Efficient TRaffic Engineering (ENTRE)* heuristic that follows the previous model and applies an online distributed *Traffic Engineering* approach that jointly balances load and energy consumption in real-time, responding to actual traffic demands. The main contribution of our approach is to provide dynamic and lightweight management of the load and the energy consumption, avoiding in this way "resource gluttony" in the network. Each ingress-egress node pair $i$ measures every $t_m$ seconds a change in the fraction of traffic ($\Delta x_{ip}$) sent along path $p$. Furthermore, *ENTRE* measures the energy "distance" ($\Delta E_{ip}$) of path $p$ from the average energy consumption of the paths between ingress-egress node pair *IE*:

$$\Delta x_{ip} = \left(\overline{U}_i - U_{ip}\right) \frac{r_{ip}}{\sum_{k \in P_i} r_{ik}}, \quad \text{when} \quad U_{ip} > U_{\min}$$

$$\Delta E_{ip} = \left(\overline{E}_i - E_{ip}\right), \quad \text{when} \quad E_{ip} > E_{\min}$$

where:

$$r_{ip} = \frac{B_{p t_m}}{t_m}, \forall p \in P_i, \forall i \in IE$$

$$\overline{U}_i = \frac{\sum_{p \in P_i} r_{ip} U_{ip}}{\sum_{k \in P_i} r_{ik}}, \forall p \in P_i, \forall i \in IE$$

$$\overline{E}_i = \frac{\sum_{p \in P_i} r_{ip} E_{ip}}{\sum_{k \in P_i} r_{ik}}, \forall p \in P_i, \forall i \in IE$$

In case that $\Delta x_{ip} > 0$ ($p$ is underutilized), the fraction of traffic sent along path $p$ must be increased by $\Delta x_{ip}$. Contrary, in case that $\Delta x_{ip} < 0$ ($p$ is over-utilized), the fraction of traffic sent along path $p$ must be decreased by $\Delta x_{ip}$. It is obvious that a possible increase in the fraction of traffic sent along a path will lead to energy consumption increase (the increased maximum utilization of that path will affect the energy consumption). Since one of the main objectives of our energy-aware approach is to keep also the maximum energy consumption in low levels, we apply the same policy for improving energy consumption ($\Delta E_{ip}>0$: Energy consumption could be increased in $p$, $\Delta E_{ip}<0$: Energy consumption must be decreased in $p$). ENTRE combines the previous policies, balances the traffic every $t_m$ seconds and jointly keeps the maximum link utilization and the path energy consumption as low as possible. We summarize the basic rules in our heuristic mechanism:

1. **IF**: $\Delta x_{ip}>0$ and $\Delta E_{ip}>0$ **DO**: Apply $\Delta x_{ip}$ to $p$
2. **IF**: $\Delta x_{ip}<0$ and $\Delta E_{ip}<0$ **DO**: Apply $\Delta x_{ip}$ to $p$
3. **IF**: $\Delta x_{ip}>0$ and $\Delta E_{ip}<0$
   a. **IF**: $\bar{E}_i - E_{ip} > T_E$ **DO**: Exclude $p$ from the routing table and turn the corresponding links into sleeping mode. Traffic is proportionally provisioned to the remaining paths.
   b. **ELSE DO**: Nothing
4. **IF**: $\Delta x_{ip}<0$ and $\Delta E_{ip}>0$
   a. **IF**: $\bar{U}_i - U_{ip} > T_U$ **DO**: Apply $\Delta x_{ip}$ to $p$
   b. **ELSE DO**: Nothing

It is true that paths with higher minimum capacity need more traffic to achieve the same utilization as smaller capacity paths. This is the main reason why $\Delta x_{ip}$ is normalized by the rates. This also makes the change in a path's traffic proportional to its current traffic share.

*Results:* We present the evaluation study of the proposed heuristic scheme compared to the optimal solutions. We consider a network topology where four ingress nodes send traffic to four egress nodes (20 routers in total). We are using OMNET++ to simulate ENTRE and IBM ILOG CPLEX Optimizer to find the optimal solutions. Fig 1*a* depicts the network throughput while the number of disjoint paths grows and Fig 1*b* depicts the energy consumption while the achieved network throughput grows (ENTRE is compared to OSPF [1]). Lastly, Table 2 presents the performance of ENTRE, in different scenarios, compared to the optimal energy saving in the network.

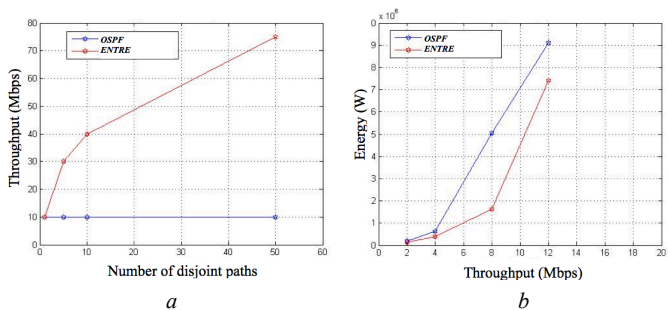

**Fig. 1** *Network throughput and energy consumption.*
*a* Throughput (OSPF vs. our approach).
*b* Energy consumption (OSPF vs. our approach).

**Table 2:** ENTRE performance

| Optimal energy saving | ENTRE energy saving | Percentage of "sleeping" links | Percentage of routes excluded | Average iterations till convergence |
|---|---|---|---|---|
| 15% | 13% | 11% | 7% | 4 |
| 26% | 23% | 24% | 18% | 7 |
| 34% | 30% | 29% | 24% | 9 |
| 43% | 38% | 41% | 31% | 12 |
| 59% | 52% | 54% | 41% | 15 |

*Conclusion:* The joint modeling of balanced and energy-efficient network operation inspired the design of a heuristic approach that tries to meet the requirements of the future core networks. The simulation results show that the proposed approach tends to behave like an optimal load balancer in the network, influenced by the minimization of the energy consumption. ENTRE converges after a small number of iterations, proving in this way it's lightweight operation.

G. Athanasiou (*Automatic Control Lab, School of Electrical Engineering, KTH Royal Institute of Technology, Osquldas vag 10, 100 44 Stockholm*)

E-mail: georgioa@kth.se